\documentclass{Rinton-P9x6}

\def\nue{\nu_{e}}
\def\num{\nu_{\mu}}
\def\nut{\nu_{\tau}}
\def\nus{\nu_{s}}
\def\nmnt{$\nu_{\mu}\leftrightarrow\nu_{\tau}$~}
\def\nmns{$\nu_{\mu}\leftrightarrow\nu_{s}$~}
\def\nenm{$\nu_{e}\leftrightarrow\nu_{\mu}$~}
\def\lsim{\lower.7ex\hbox{${\buildrel < \over \sim}$}}
\def\gsim{\lower.7ex\hbox{${\buildrel > \over \sim}$}}

\begin{document}

\title{Recent results from the Super-Kamiokande experiment\footnote{Talk
at Cairo International Conference on High Energy Physics (CICHEP2001),
Cairo, Egypt, January~9-14,~2001.}}

\author{Yuichi Oyama\\
for Super-Kamiokande collaboration}

\address{Institute of Particles and Nuclear Studies,\\
High Energy Accelerator Research Organization (KEK)\\
Oho 1-1, Tsukuba, Ibaraki 305-0801, Japan\\
E-mail:~~yuichi.oyama@kek.jp\\}

\maketitle

\abstracts{
New physics results from the Super-Kamiokande experiment in 2000 are
presented.
}

\section{Introduction}
It is about 5 years from the start of the Super-Kamiokande (SK)
experiment, and most of the physics results produced in the early stage were
published\cite{SK1,SK2,SK3,SK4,SK5,SK6,SK7,SK8,SK9,SK10,SK11,SK12}
by 1999.
Therefore, this report concentrates on two new results
obtained in 2000: one is a ``\nmnt or \nmns''
analysis of the  atmospheric neutrino data;\cite{SK13}
the other is a determination of the solar neutrino
oscillation parameters among 4 possible solutions.\cite{SK14,SK15}

\section{``\nmnt or \nmns'' analysis}

In previous papers,\cite{SK1,SK2,SK4,SK8,SK12}
the SK group reported evidence for neutrino oscillations
with atmospheric neutrinos.
The data show a strong zenith angle- dependent deficit for muon neutrinos,
and no such deficit for electron neutrinos.
Anomalous zenith-angle distributions can be well explained
by neutrino oscillation of $\num$ to some neutrinos
other than $\nue$.
The present best-fit oscillation parameters by SK are
$(\Delta m^{2},\sin^{2}2\theta) = (3.2\times 10^{-3}{\rm eV^{2}},1)$.

The next step is to identify the oscillation partner of $\num$.
The most plausible scenario is that $\num$ oscillate to $\nut$.
In this case, most of the $\nut$ can interact only through a neutral current,
because the neutrino energy is below the 3.4GeV threshold of
the charged current interaction.
An alternative scenario
is oscillation with sterile neutrinos ($\nus$),
defined as neutrinos which do not interact through a charged-current
or neutral-current interactions.

A total of 1144~days of atmospheric neutrino data,
which corresponds to 70.5~kt$\cdot$yr of exposure, were used to distinguish
between \nmnt and \nmns.
During this period, 9178 fully contained (FC) events
and 665 partially contained (PC) events were collected.
The particle type of each Cherenkov ring is identified
as e-like or $\mu$-like.\cite{Kasuga}
In the current FC samples, there are 3107 single-ring e-like events,
2988 single-ring $\mu$-like events, and 3083 multi-ring events.
In addition, 1269 upward through-going muons 
(UTM) produced by atmospheric neutrino interactions in the surrounding
rock were also employed in the oscillation analysis.

The distinction between \nmnt and \nmns utilizes 
the difference in neutral-current interactions and
matter effects in the Earth.

\subsection{Neutral-current interactions}

A $\nus$ does not interact with matter even through neutral current,
while a $\nut$ interacts through the same neutral current
as the original $\num$. Therefore, 
for the \nmns oscillation, one should observe fewer neutral current
events than for the \nmnt oscillation.
This difference is more significant for upward-going neutrinos
because of their long travel distances.

In order to obtain a sample enhanced by neutral current events,
we applied the following selection criteria for FC events:
(1)vertex within the fiducial volume; (2)multiple Cherenkov rings;
(3)particle identification of the brightest
ring is e-like; and (4)visible energy greater than 400MeV.
The first criterion provides a contained event sample,
and the second and third criteria serve to enrich the neutral current
event fraction.
The fourth criterion helps to obtain good angular
correlation between the incident neutrino and the reconstructed direction,
defined as the pulse-height weighted sum of the ring direction.
The remaining number of events from this selection is 1531, and about 29\%
of them are expected to be neutral-current interactions
in the case of a null oscillation.

In figure~\ref{fig1}(a),
the zenith-angle distribution is plotted together with the expectations
for the \nmnt and \nmns oscillation with the oscillation parameters,
$(\Delta m^{2},\sin^{2}2\theta) = (3.2\times 10^{-3}{\rm eV^{2}},1)$.
The data are consistent with the \nmnt oscillation,
while the data differ from the prediction for the \nmns oscillation
by 2.4 standard deviations.

\subsection{Matter effects in the Earth}

The interaction of neutrinos with matter\cite{matter}
leads to a difference in
the oscillation probability. Since the coherent forward scattering
of $\num$ and $\nut$ are identical,
the presence of matter in the neutrino path does
not modify the oscillation probability. In contrast, $\nus$ does not interact
with matter, even through a neutral current.
This introduces an effective potential
which modifies the mixing angle and oscillation length.
If the oscillation parameter region suggested by SK is assumed,
the oscillation probability
of upward-going neutrinos with energy greater than $\sim$15~GeV would
be suppressed by matter effects in the \nmns case.
In consequence, the number of upward-going events for the \nmns oscillation
should be larger than that for the \nmnt oscillation.

\begin{figure}[t!]
\centerline{
\epsfig{file=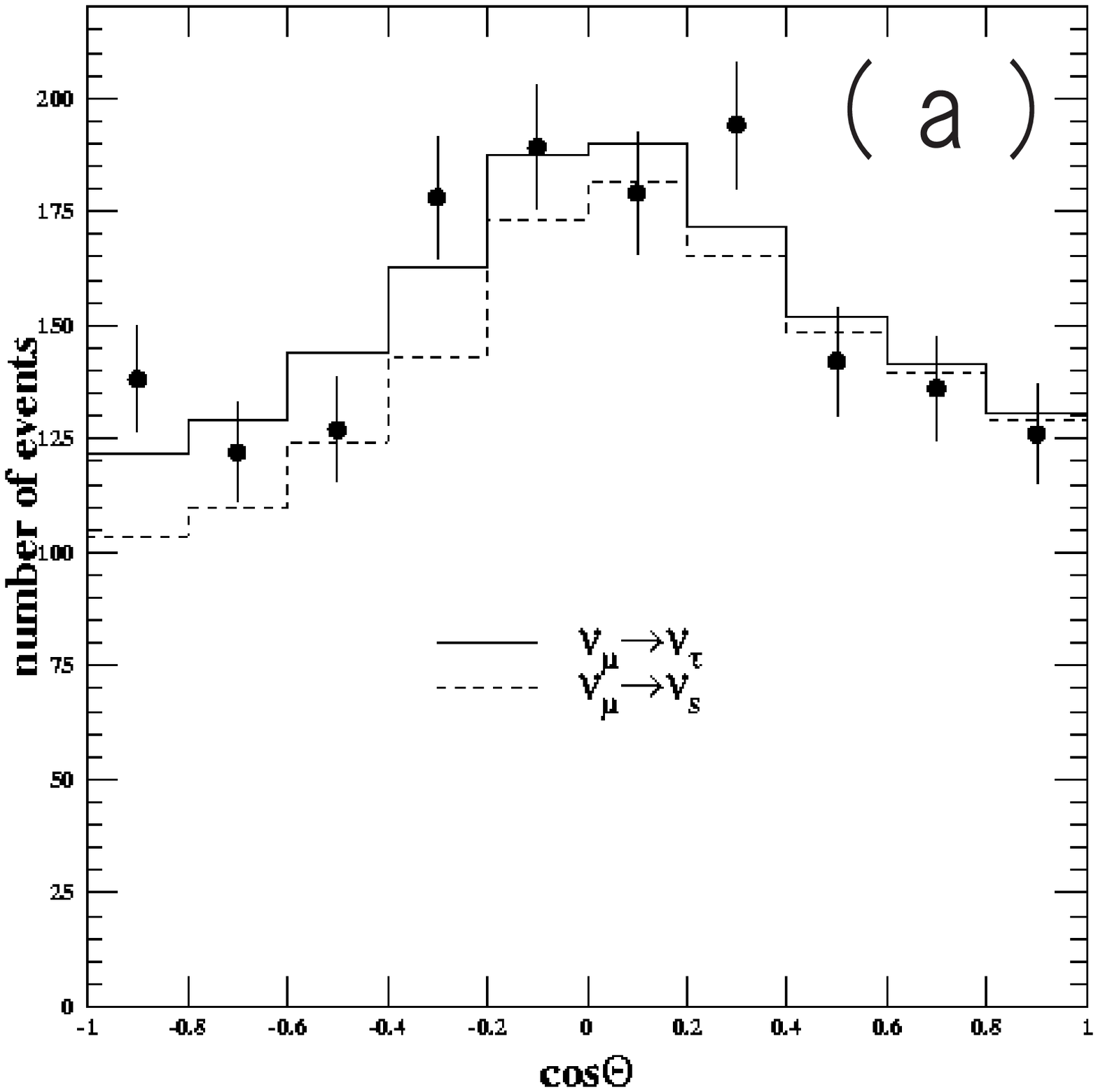,height=4.0cm}
\hskip 0.3cm
\epsfig{file=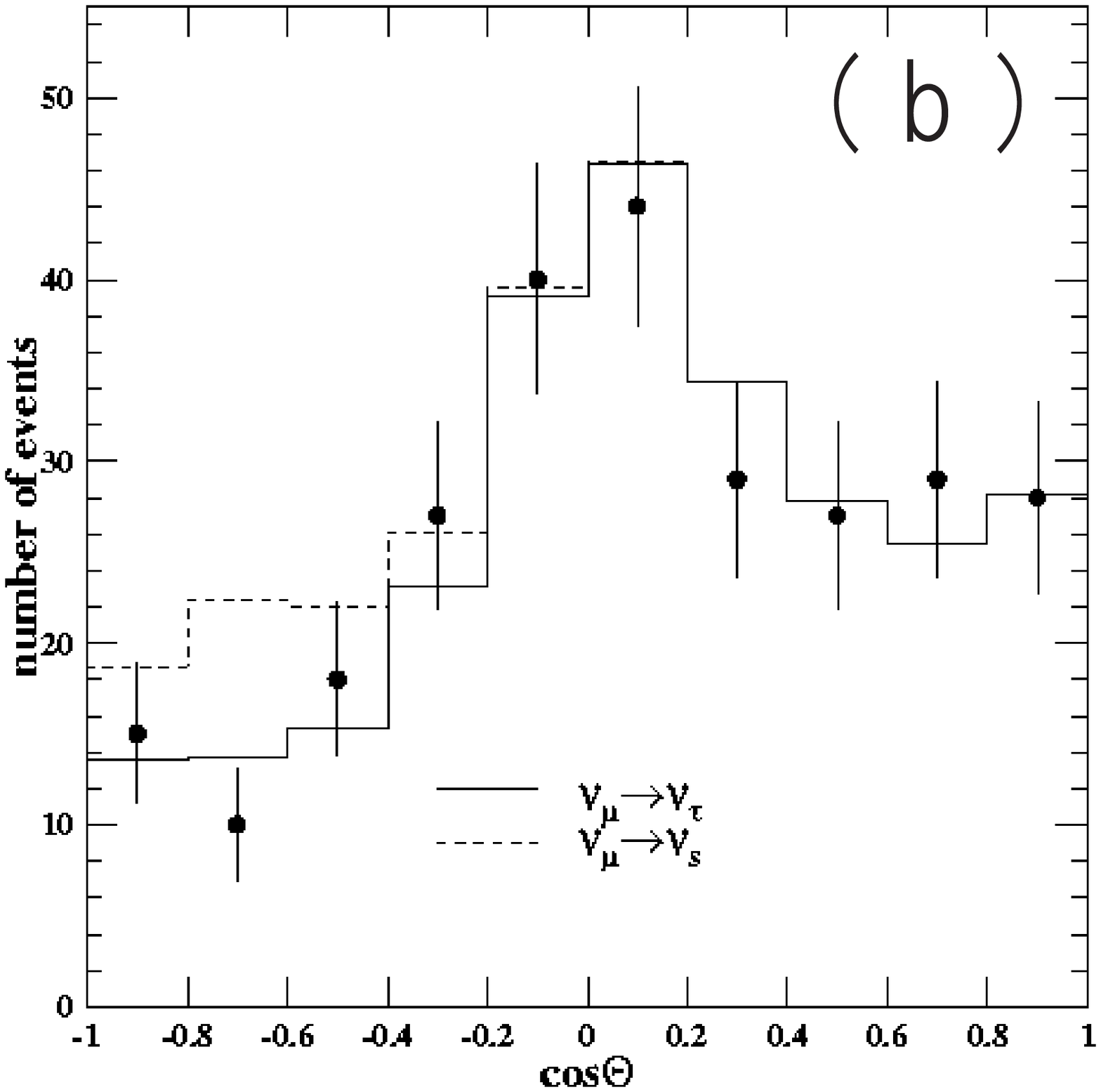,height=4.0cm}
\hskip 0.3cm
\epsfig{file=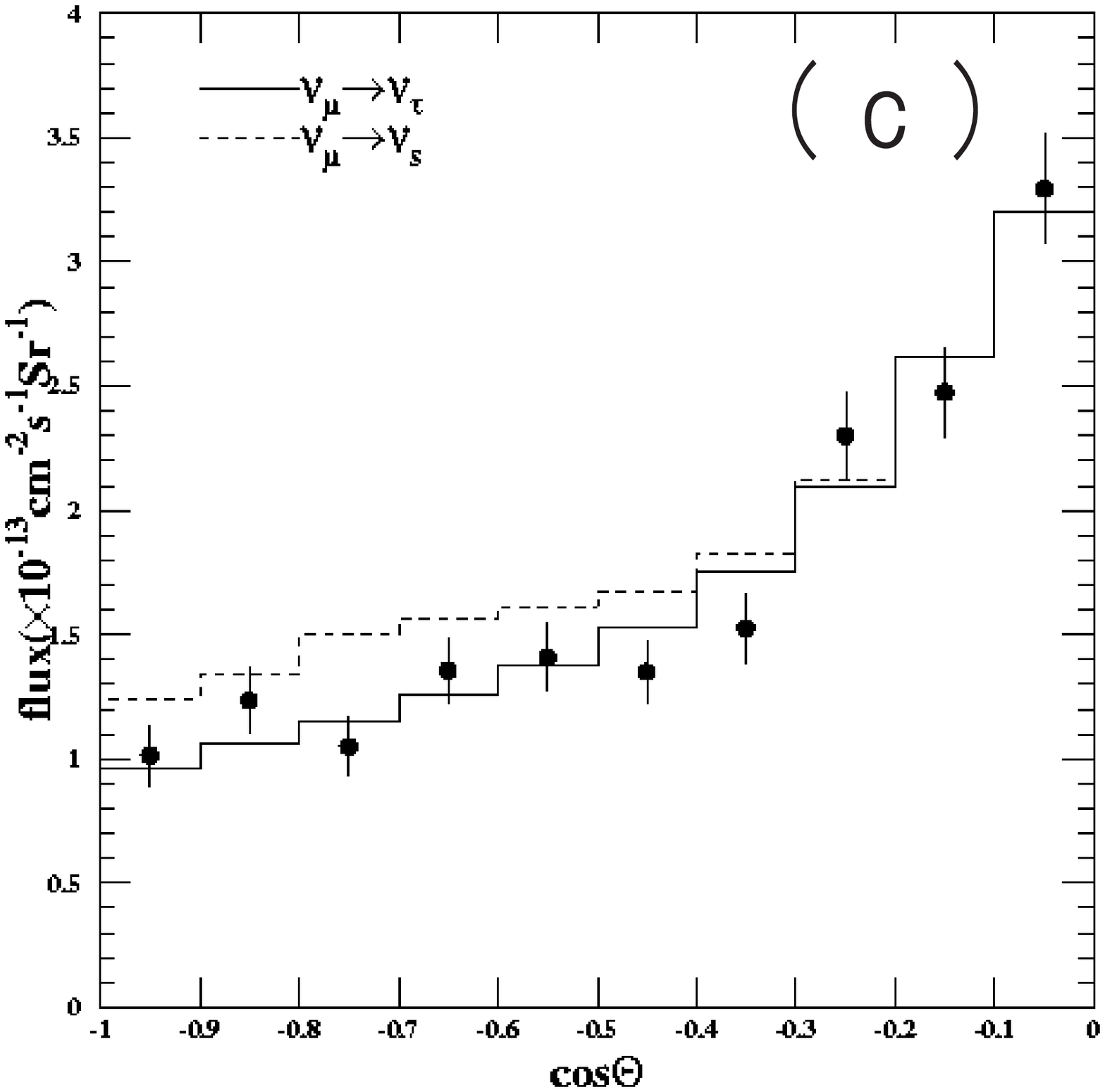,height=4.0cm}
}
\caption{
Zenith-angle distributions of atmospheric neutrino events
satisfying the cuts described in the text:
(a)multi-ring sample, (b)partially contained sample, and
upward through-going muon sample.
The filled circles indicate the data with statistical errors.
The solid line indicates the prediction for \nmnt, and
the dashed for \nmns, with
$(\Delta m^{2},\sin^{2}2\theta) = (3.2\times 10^{-3}{\rm eV^{2}},1)$.
The two predictions are independently normalized to the
number of downward-going events for (a) and (b), and
the number of horizontal events for (c).
}
\label{fig1}
\end{figure}

Two different data samples are employed to examine the matter effect.
One is high-energy PC events.\cite{SK2}
They are estimated to be 97\% pure
$\num$ charged current with a mean neutrino energy of 10 GeV.  In
order to select higher energy $\num$ events, which are more
sensitive to matter effects, we additionally require visible energy
greater than 5~GeV. The typical energy of the parent
atmospheric neutrino is estimated to be 20~GeV. After cuts were made upon the current
data sample we found 267 events. 
Figure \ref{fig1}(b) shows the
zenith-angle distribution of these events and predictions for
$(\Delta m^{2},\sin^{2}2\theta) = (3.2\times 10^{-3}{\rm eV^{2}},1)$.
The results are consistent with the \nmnt oscillation,
whereas it differs from
the \nmns oscillation by 2.3 standard deviations.

The other event sample was upward
through-going muon events.\cite{SK8}
Because the typical energy of the UTM parent neutrino is
approximately 100~GeV, matter-effect suppression should appear most
prominently in this data set.
Figure \ref{fig1}(c) shows the zenith-angle distribution of these events
with predictions. 
For $(\Delta m^{2},\sin^{2}2\theta) = (3.2\times 10^{-3}{\rm eV^{2}},1)$,
the data are consistent with the \nmnt oscillation, while the \nmns
oscillation differs from the data by 2.9 standard deviations.

\smallskip

Three independent event samples show that the \nmnt oscillation is favored.
If three results are combined, the \nmns oscillation can be excluded with
99\% C.L.

\section{Determination of oscillation parameters for solar neutrinos}

Solar-neutrino fluxes have been measured by 3 different types of
experiments: radio chemical measurements with chlorine and gallium,
and water Cherenkov experiments.
The absolute solar neutrino fluxes from these experiments have been significantly
smaller than the prediction of the Standard Solar Model (SSM).\cite{bp2000,ssm}
This ``solar neutrino problem'' is generally believed to be due to
neutrino flavor oscillation between electron neutrinos and other spices.

\begin{figure}[t!]
\centerline{
\epsfig{file=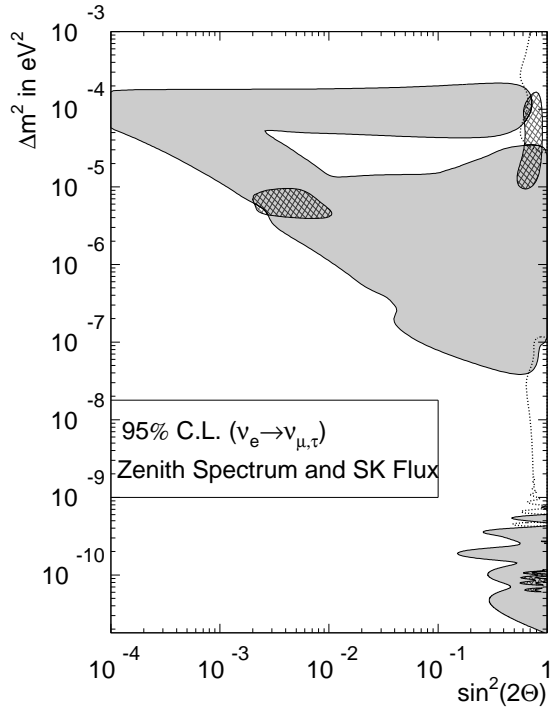,height=10.cm}
}
\caption{
Constraints for two-flavor oscillation, \nenm or $\nut$.
The three hatched areas show 95\% allowed region from a combined
fit of the absolute fluxes from Homestake, SAGE, GALLEX and SK.
They are, namely, LMA solution (top-right), SMA solution (top-left)
and Just-so solution (bottom-right).
The shaded areas show 95\% C.L. excluded regions from
zenith angle and spectrum analysis by SK.
Allowed regions with 95\% C.L. by absolute flux, zenith angle and
spectrum in SK are also shown by the dotted lines. 
}
\label{fig2}
\end{figure}

The most recent results in SK are based on 1258~days of
data from May~31,~1996 to Oct.~6,~2000.
The detection threshold for the recoiled electrons is 5~MeV.
The updated solar neutrino flux is
$2.32\pm 0.03({\rm stat.})\hbox{${{+0.08}\atop{-0.07}}$}({\rm sys.})
\times 10^6~{\rm cm}^{-2} {\rm s}^{-1}$,
which corresponds to 
$45.1\pm 0.5({\rm stat.})\hbox{${{+1.6}\atop{-1.4}}$}({\rm sys.})\%$
of the theoretical expectation(BP2000).

To obtain constraints on the oscillation parameters,
the fluxes measured by SK are combined with results from other
experiments.
The other solar-neutrino fluxes used in the analysis are;
$2.56\pm 0.16\pm 0.16$ SNU (Homestake\cite{Homestake}) for $E_{th}=0.814$MeV,
75.4\hbox{${{+7.8}\atop{-7.4}}$}SNU (SAGE\cite{Sage})
and 74\hbox{${{+6.7}\atop{-6.8}}$}SNU
(GALLEX\cite{Gallex}) for $E_{th}=0.233$MeV.
An average of SAGE and GALLEX, $74.8\pm 5.1$ SNU, is used as
the numbers for gallium experiments.
The constraints on the neutrino oscillation parameters,
$(\Delta m^{2},\sin^{2}2\theta)$ for the \nenm or $\nut$ oscillation
are shown by the shaded region in Figure \ref{fig2}.
At present, three possible parameter regions remain;
they are the LMA (Large Mixing Angle) solution,
the SMA (Small Mixing Angle) solution, and the Just-so solution.
The LOW solution, which was allowed in past analysis, 
is now rejected with 95\% C.L. because of updates of the
neutrino flux data.

The next step is to determine the oscillation parameters among
three possible regions.
For this purpose, the energy spectrum and solar zenith-angle distribution
of the SK data are employed because
the LMA solutions show a solar zenith angle variation,
while the SMA and the Just-so solutions show a distortion of the neutrino
energy spectrum.
It should be stressed that
these analyses are only possible by SK because
the SK observation is real-time,
directional, and energy sensitive.

\begin{figure}[t!]
\centerline{
\epsfig{file=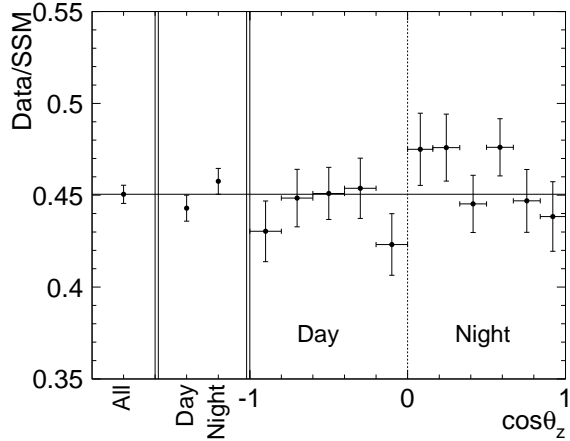,height=6.cm}
}
\caption{
Solar zenith angle dependence of the solar-neutrino flux.
The solar zenith angle, $\theta_{z}$, is defined as the angle
between vertical axis at SK and the vector from the Sun to the
Earth. The error bars show statistical errors.
The width of night-time bins was chosen to separate solar neutrinos
that pass through the Earth's dense core
($\cos\theta_{z} \geq 0.84$)
from those that pass through the mantle
($0 < \cos\theta_{z} < 0.84$).
The horizontal line shows the flux for all data
}
\label{fig3}
\end{figure}

Figure~\ref{fig3} shows the solar-neutrino flux as a function of the solar
zenith angle, $\theta_{z}$. 
The LMA solution predicts a non-zero difference between
$\Phi_{d}$ and $\Phi_{n}$ due to the matter effect
in the Earth's mantle and core,
where the day-time flux $\Phi_{d}$ is defined as
the flux of events when $\cos\theta_{z} \leq 0$, while the night-time
flux, $\Phi_{n}$, is that when $\cos\theta_{z} > 0$. 
The measured fluxes are
$\Phi_{d}=2.28\pm 0.04({\rm stat.})\hbox{${{+0.08}\atop{-0.07}}$}({\rm sys.})
\times 10^6~{\rm cm}^{-2} {\rm s}^{-1}$ and
$\Phi_{n}=2.36\pm 0.04({\rm stat.})\hbox{${{+0.08}\atop{-0.07}}$}({\rm sys.})
\times 10^6~{\rm cm}^{-2} {\rm s}^{-1}$.
The degree of this difference is given by the day-night
asymmetry, defined as
${\cal A}=2(\Phi_{n}-\Phi_{d})/(\Phi_{n}+\Phi_{d})$.
We find ${\cal A}=0.033\pm 0.022({\rm stat.})\hbox{${{+0.013}\atop{-0.012}}$}
({\rm sys.})$.
If systematic errors are included,
it is a 1.3$\sigma$ deviation from zero asymmetry.

\begin{figure}[t!]
\centerline{
\epsfig{file=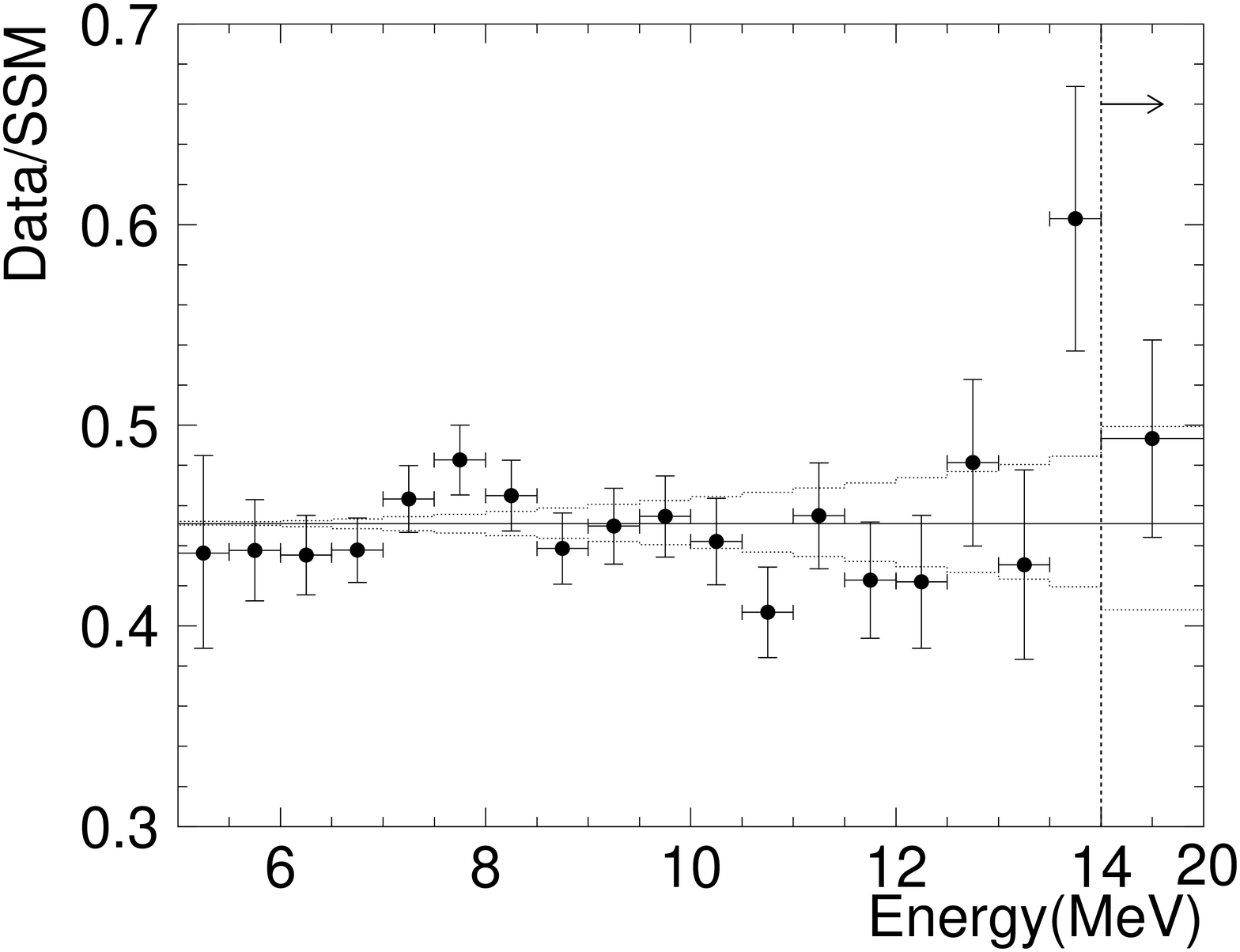,height=6.cm}
}
\caption{
Measured $^{8}$B + ${\it hep}$ solar-neutrino spectrum
relative to that of Ortiz ${\it et~al.}$
normalized to BP2000.
The data from 14~MeV to 20~MeV are combined into a single bin.
The horizontal solid lines show the measured total flux,
while the dotted band around this line indicates
the energy-correlated uncertainty.
The error bars show the statistical
and energy-uncorrelated errors added in quadrature.
}
\label{fig4}
\end{figure}

The energy spectrum of recoiled electrons relative to a
theoretical expectation is shown Fig.\ref{fig4}. The expectation is based
on the energy spectrum shape by Ortiz {\it et al.}\cite{Ortiz},
and an absolute normalization by BP2000.\cite{bp2000}
The energy spectrum is consistent with a flat distribution.
A fit to the undistorted energy spectrum gives $\chi^{2}=19.0/18.$
This corresponds to the 39\% confidence level for the flat distribution.
The absence of distortion is favored by the LMA solution.

The results from the day-night difference and energy spectrum
is combined, and constraints on the $\Delta m^{2} - \sin^{2}2\theta$
plane are obtained. The detailed calculation method
is reported in Ref.15, and is not discussed here.
The 95\% C.L. excluded regions are shown in Figure~\ref{fig2}.
The Just-so solution is completely rejected with 95\%,
and most of the SMA solution is also rejected.
Although a small region still remains in the SMA region,
the LMA solution is the most probable solution for the
solar-neutrino problem.

\end{document}